# Multifractal test for nonlinear changes in time series

Damian G. Kelty-Stephen[1], Elizabeth Lane[1], and Madhur Mangalam[2]

[1]Department of Psychology, Grinnell College, Grinnell, IA, USA

[2]Department of Physical Therapy, Movement and Rehabilitation Sciences, Northeastern University, Boston, MA, USA

ORCIDs:

Damian G. Kelty-Stephen (0000-0001-7332-8486)

Madhur Mangalam (0000-0001-6369-0414)

E-mails: keltysda@grinnell.eud;  elizabeth.lane96@gmail.com;  m.manglam@northeastern.edu

**Abstract**

The creativity and emergence of biological and psychological behavior are nonlinear. However, that does not necessarily mean only that the measurements of the behaviors are curvilinear. Furthermore, the linear model might fail to reduce these measurements to a sum of independent random factors, implying nonlinear changes over time. The present work reviews some of the concepts implicated in linear changes over time and details the mathematical steps involved. It introduces multifractality as a mathematical framework helpful in determining whether and to what degree the measured time series exhibits nonlinear changes over time. The mathematical steps include multifractal analysis and surrogate data production for resolving when multifractality entails nonlinear changes over time. Ultimately, when measurements fail to fit the structures of the traditional linear model, multifractal modeling gives us the means to make those nonlinear excursions explicit and perhaps permit the development of theory that draws on both linear and nonlinear processes.

## Introduction

The analysis of linear or nonlinear changes over time evaluates whether or not the measured time series can be effectively modeled as a sum of independent random factors, or the sum of the measurements (Fig. 1*a*). Nonlinear changes can mean that the time series is not well-modeled as merely a sum. The question of whether it is a sum is a foundational mathematical issue and goes deeper than the relatively superficial question of whether changes over time 'look' linear to the eye. Linear changes over time can include curvilinearity; some nonlinear trajectories are perfectly compatible with modeling the time series as a sum. For instance, time-series with peaks and valleys can invite a polynomial model. Polynomial models include linear effects of time and several powers of the time (e.g., quadratic and cubic for second and third powers). Critically, polynomials are linear in the growth parameters, or sums after all—sums of integer exponents of time (e.g., linear growth is proportional to 'time' which is just time raised to the power of 1, i.e., time$^1$,' but quadratic profiles are just the sum of 'time$^1$' with 'time squared' which is just time raised to the power of 2, i.e., 'time$^2$,' and cubic profiles are 'time$^3$ + time$^2$ + time$^1$' and so forth). Choosing nonlinearity as a failure to reduce to a sum or a linear model is mathematically deeper than just eyeballing the plots. That is, a curvy plot of data points across time may still reduce to a sum of independent factors, but 'seeming linear' does not guarantee that it is actually linear. The linear model (i.e., summing parts) can make very many seemingly linear changes with time, but it can also produce very many curvilinear profiles. So, it is important to distinguish between 'seeming linear' and being a linear sum.

Multifractality can provide us deeper insights into the failure of a time series' nonlinearity to reduce to a sum of independent random factors (Fig. 1*b*). It is quickly becoming clear that the failure of a time series's nonlinearity to reduce to a sum is important to psychology. Beyond resemblance to old Gestalt wisdom that wholes differ from sums of parts, estimates of multifractality have predicted outcomes in executive function, perception, or cognition, such as in reaction time (Ihlen and Vereijken 2010), gaze displacements (Kelty-Stephen and Mirman 2013), word reading times (Booth et al. 2018), speech audio waveforms (Hasselman 2015; Ward and Kelty-Stephen 2018), rhythmic finger tapping (Bell et al. 2019), gestural movements during the conversation (Ashenfelter et al. 2009), electroencephalography (Kardan et al. 2020a), and functional magnetic resonance imaging (Kardan et al. 2020b). Our purpose is not to review the empirical meaning of multifractality in psychological terms; this question may not even be answerable in full at present. Multifractality is the logical consequence of processes that enlist interactions across scales (Ihlen and Vereijken 2010), suggesting that it is essential to processes unfolding at many rates, such as Gottlieb's (2002) probabilistic epigenesis. However, the truth would be better served with a broader set of scholars exploring the role of multifractality in psychological processes. So, our purpose is to make the method more accessible.

This tutorial introduces multifractality as a mathematical framework helpful in determining whether and to what degree a time series exhibits nonlinear changes over time. It is by no means the first to introduce multifractality; prior entries to the multifractal-tutorial literature have sometimes taken a more conceptual perspective, introducing nonlinearity across time as an interaction across multiple scales (Kelty-Stephen et al. 2013). Other tutorials have kept closer to detailing algorithmic steps through the use of computational codes (Ihlen 2012). The present work aims to tread a middle ground, reviewing some of the concepts implicated in linear and nonlinear changes over time and detailing the mathematical steps involved. These mathematical steps include multifractal analysis and surrogate data production for resolving when multifractality entails nonlinear changes over time.

## What is multifractality, what factors impact changes in multifractality over time, and how is multifractal analysis useful?

Multifractality, a modeling framework developed in its current form about fifty years ago (Mandelbrot 1976; Halsey et al. 1986), is primarily a statistical measurement of heterogeneity in how systems change across time. All mathematical frameworks work by encoding more variability into a symbolic and logical structure. Multifractality is no exception. What multifractality encodes is the heterogeneity, and it encodes this heterogeneity as a range—maximum minus minimum—of fractional exponents. These exponents represent the power-law growth of proportion and time scale. This relationship between proportion and scale is a pervading question for any time we observe a changing system we want to understand: all time-varying processes vary with time, and we are constantly dealing with the issue that a smaller sample of the whole process tells us something but not everything about that whole process. So, the important question is, "how long do we have to look before we see a representative sample of the time-varying process?" The proportion of the process we can see will increase the longer we look, and that proportion increases nonlinearly, with the proportion increasing as a power function (also called a 'power law') of scale, with the proportion increasing with scale$^{alpha}$. Multifractality becomes useful when there is not simply one alpha value when, for various reasons outlined below, there may be many. That means that multifractality can help us understand how and why our samples of observations may align with the broader structure of the process. In summary, multifractality encodes heterogeneity as the range—maximum minus minimum—of fractional exponents that govern the power laws relating the observed proportions of heterogeneous changes to a specific timescale (i.e., how a change in measurement relates to a proportional change in time).

Multifractality arose from a long history of scientific curiosity about how fluid processes generate complex patterns (Richardson 1926; Turing 1952) and remains one of the leading ways to model fluid, nonlinear processes—as initially intended in hydrodynamics (Schertzer and Lovejoy 2004) and more recently as a framework for understanding the fluid-structure of perception, action, and cognition (Dixon et al. 2012; Kelty-Stephen 2017; Kelty-Stephen et al. 2021). This short answer is only clear from a specific background in time-series analysis. So, in this essay, we unpack both what multifractality is and why it is helpful for quantifying nonlinear changes in time series, with specific examples from perception, action, and cognition.

**What is the relationship between multifractality, mean, and standard deviation?**

Multifractality sits apart from the more familiar descriptive statistics like mean and standard deviation—indeed, it does not replace mean and standard deviation. A good reason for the widespread use of mean and standard deviation is that they support a wide range of inferential methods to test the effects of many types of hypothesized causes. However, when our hypotheses about causes begin to probe the issue of changes over time, mean and standard deviation no longer suffice. Mean and standard deviation never fail to be helpful or necessary to statistical reasoning but fail to cover more complex relationships that evolve continually. Therefore, the use of mean and standard deviation alone does not test hypotheses about how systems change [nonlinearly] over time.

**Why are mean and standard deviation not enough to model how a system changes over time?**

Means and standard deviations accord with the linear assumptions yet reflect inadequacies for cases that do not fulfill the assumption of nonlinearity (Mandic et al. 2008). The linear assumption comes with the expectation that means and standard deviations are stable over time. Changes with time require us to acknowledge that the linear model has three defining features: 1) mean, 2) standard deviation 3) specification of the linear autocorrelation (or equivalently the amplitude spectrum of the Fourier transform). The linear autocorrelation describes how a given time-varying process correlates with past behavior (e.g., how behavior over the current month resembles behavior over the past month). It expresses how correlated current behavior is to past behavior, at many different lags (or periods being considered), and contains a coefficient for each 'lag.' For instance, the autocorrelation's coefficient for the lag of one would indicate each value's average similarity to the value just one unit of time beforehand. The coefficient for the lag of two would indicate each value's average similarity to the value two units of time beforehand. There are only $N$-1 coefficients, where $N$ is the length of the time series, because the linear autocorrelation compares present to past, over all of the available lags across time, and $N$-1 would be the largest possible lag of time to compare two different values of a time series.

The Fourier transform attempts to describe a time series as a set of sinusoidal waves, that is, as a set of regular, periodic oscillations. This description uses the fact that sinusoidal waves can vary both in frequency (or length of time taken for each cycle) and amplitude (i.e., the size of the oscillations). The Fourier transform breaks a time series into component sinusoidal waves of many different frequencies (or wavelengths) and then describes the time series by estimating the amplitude of sinusoidal variation at each frequency. Change can be large or small, and change can also be fast or slow. The Fourier transform examines change at many different speeds, and it estimates how large the change is at each speed. The list of amplitudes for each corresponding frequency for all possible sinusoidal waves is called the 'amplitude spectrum.' An interesting feature here is a one-to-one relationship between the linear autocorrelation and the Fourier spectrum (Wiener 1964). That is to say, the autocorrelation across different lags carries similar information as the oscillatory amplitudes across different frequencies.

The linear autocorrelation exists even when it is not explicitly described. Unless otherwise stated, its value is always zero. In the pedagogical example of a fair coin toss, it is assumed as part of the 'fairness' of the coin that the likelihood of heads or tails does not change over time. So, the fair coin is almost always assumed to have zero linear autocorrelation (i.e., random chance). Of course, instead of explaining 'fairness' of a coin flip in terms of linear autocorrelation, each coin flip is said to be independent of the past coin flips. However, this independence implies an absence of temporal structure. The linear autocorrelation offers the linear model a very subtle way to model how the present events might be products of specific events in the past. Notwithstanding, changes over time can be entirely linear, and without evidence to the contrary, nonlinearity is not necessary to modeling. It only becomes necessary to modeling as changes over time fail to follow linear patterns. Similarly, the Fourier spectrum is as flat and uniform as the zero autocorrelation in the fair coin example. The fair coin entails what is called 'white noise,' which exhibits equal amplitudes for oscillations across all possible frequencies.

**What is entirely linear change in our dependent measures over time?**

A linear model applied to a developing process assumes that it changes similarly over time in the beginning, in the middle, and at the end of that process. This symmetry across time that linearity assumes appears most clearly in the sinusoidal waves that the Fourier transform uses to decompose a time series.

Ironically, this elegant model for changes over time is locked into repeating the same changes over time. The Fourier-transform's use of sinusoidal waves to model changes over time reflects the underlying premise that all changes reverse (e.g., all changes over time balance out), or more widely known as "regression to the mean," the idea that what changes over time will hover around the mean (e.g., what goes up must come back down). Whether we use the linear autocorrelation or the Fourier amplitude spectrum to quantify changes over time, the linear model necessarily expects a relationship between the past and the present that does not itself change.

**When should we use something other than the linear model to understand changes in a system over time?**

It is worth consulting nonlinear methods anytime the linear model fails to exhaust the empirically observed variability. However, the linear model is an exceptionally compact and effective statistical framework. It is important to note that multifractality remains unfamiliar to some audiences because the statistical literature often frames nonlinear problems with linear solutions. Let us imagine that a developing process exhibits different changes over time at the beginning versus at the end. This issue can be resolved by finding a proposed breakpoint where a sudden event that brought about an abrupt change and behavior followed an entirely different pattern after that event. For instance, we might make the same set of footpaths on the way to work every day and back home every day, but an earthquake could suddenly damage or change the footpaths with trees or structures along the way. After the earthquake, we might take a radically different set of footpaths to get to and from work every day. So, if it were our job to determine the best fitting model of our footpaths over very many days, it might make a lot of sense to just fit one model for the pre-earthquake footsteps and a separate model of post-earthquake footsteps.

Although it is sensible to model our time series differently around supposed breakpoints, this strategy would depart from the linear model that appeals to a scientist capable of starting a new model. The creation of a new model can be systematized by fitting long-term predictors for different-sized bins. For instance, predicting daily spending behavior might involve looking at short-term predictors such as individual previous days' spending behavior. However, quite apart from these short-term, day-to-day changes, longer-term trends may be more predictive. For instance, for a college student, spending behavior may differ vastly during the school year from during the summer. Summer months may offer the possibility of full-time employment, and so the effect of previous days' spending behavior may be entirely different during the longer-term period of summer versus the rest of the year. Such long-term predictors are often called 'seasonal' and suggestive of cyclic repetition, e.g., summer arrives reliably at the same point on the yearly school calendar. The challenge is that identifying those long-term predictors requires theory and intuition but also account for long-term periodicities or bins in them. Indeed, once a student graduates or leaves school, the summer's cyclic effect may disappear as they begin to work full-time all year round. At such a point, the relevant cyclic patterns useful for predicting spending behavior may change.

At least two good reasons exist to wish for an alternative to strictly linear models of changes over time. First, "changes over time" may be non-cyclically continuous; that is, changes may shift over time without any simple breakpoints. The lack of cyclicity may be necessary because a system may change without returning to the initial 'normal.' It is essential to underscore that any expectations of "regression to the mean" result from the linear assumption of temporal symmetry. The temporal symmetry of linear models means they look the same played backward in the past as they do play forward into the future (Lutkepohl 2007). However, there is no statistical guarantee that what goes up must come down as most measured systems grow, mature and decay.

The second reason to wish for an alternative to strictly linear models is the interaction among changes over time across cycles. For instance, roughly cyclic periods like the year, the month, and the day can be easily identified. However, a daily routine may vary considerably across the span of a month (e.g., around weekly or biweekly salary payments), and it may vary further across different months in a year (e.g., as holiday bonuses and time off allow various ways to behave). The necessity to account for the interaction of various differently-scaled factors over time has prompted the need for multifractal modeling. While there are hierarchical linear models that can measure short- and long-term effects, they estimate these interactions with the same expectation of consistency across time—also known as 'stationarity' (Singer and Willett 2003).

**Examples of changes over time**

Given the origins of multifractal modeling in fluid dynamics (References; References; References), an apt example to consider is daily rainfall in a given region. Daily rainfall can be measured in centimeters to examine how it changes over time. Reasons for abrupt changes include elevation, humidity, and temperature. Reasons for more sustained changes can include the seasons and movement of tectonic plates. Appearance or disappearance of currents, winds, and vegetation can also impact daily rainfall.

**Perfect stability.** The linear model aptly applies to changes in daily rainfall over time under many circumstances. The simplest measurements of daily rainfall include 1) no rainfall (i.e., perfect drought) or 2) always the same amount of rainfall (Fig. 2*a*). In both cases, the present looks perfectly like the past and the future. Such processes are temporally symmetric. They have perfectly stable means: drought entails a zero mean for the entire time series, and the same amount of rainfall reflects a stable mean for the entire time series. Of note, the standard deviation is zero in both cases.

**White noise.** Another realistic case perhaps more suitable to areas with more temperate climates would be (3) uncorrelated random variation of daily rainfall, varying according to 'white noise' (Fig. 2*b*). White noise is the statistical term for the product of many independent processes. Calling it 'white' reflects an almost poetic allusion to the fact that some of the earliest uses of the Fourier transforms involved the application to electromagnetic radiation (i.e., light), and some models of white light have indicated a broadband contribution of radiation oscillating at many visible frequencies (Loudon and Murphy Jr 1984). Therefore, white noise epitomizes the temporal symmetry characteristic of linear changes over time and the regression to the mean. A histogram of a white-noise process will approach (over long time scales) a Gaussian (or Normal) distribution with stable mean and standard deviation. This Gaussian profile of white noise is a close statistical cousin of the binomial distribution that a fair coin would generate for large samples of progressively longer sequences of coin flips (Box et al. 1986). Crucially, white noise regresses to the mean in the long run, and in the short term, it is uncorrelated in time (here, 'uncorrelated' implies no correlation between rainfall across days, weeks, or months). In this case, the average rainfall for one day, one week, or one month is as good a predictor of the next day, week, or month, respectively, as of any other day, week, or month, respectively. In other words, the average rainfall of one time period predicts future time periods—in large part because all of this sequence is statistically the same.

**Uniform seasonality.** The measurement area may have a rainy season and, therefore, a cyclical rainfall pattern (e.g., more in June and July than in April; Fig. 2*c*). So long as this rainy season begins and ends reliably with the exact dates, the linear model will produce an adequate description of this rainfall. The changes over the year would show a peak across months, but in this example, with perfect timing of the seasons, these changes across months do not change from year to year.

**Irregular seasonality.** The above examples are rare cases in practice but help illustrate the temporal symmetry of linear models. However, rainfall is often more irregular (Fig. 2*d*). Rainy seasons can come late one year or early; they can come early multiple years and late the following year. Also, wet or dry years can exist in which the rainy season varies in intensity across years, decades, and centuries. When considering these longer timescales, the Fourier transform can spread longer and longer sinusoidal waves, and similarly, the linear autocorrelation can incorporate progressively longer lags or waves. In any event, as rainfall is measured over long timescales, our linear model can be complicated with progressively more factors. However, no matter how long the timescale or added factors, the linear model's constraint is that the cyclical patterns must be regular across time.

Linear models may be perfectly valid with small errors, but due to the expected error (i.e., differences between measured and predicted) across time, no one expects the empirical record to be perfectly regular. No linear model may thus demonstrate a perfect fit. It would look bizarre for a model to predict perfectly. For instance, if our rainfall model predicted 26, 3, and 15 cm of rainfall next Monday, Tuesday, and Wednesday, we might find that the actually measured rainfall turned out to be 19, 10, and 11 cm. That would entail errors in the prediction of 7, -7, and 4 for those three days. Smaller errors beget more accurate predictions than larger errors. However, so long as those errors exhibit the lack of temporal structure as white noise, then the model is still a valid linear model. That is, it is not required for measurements of the developing process to have the same mean and variance. Only prediction errors must have a zero mean, stable variance, and no correlation across days, that is, non-zero autocorrelation coefficients overall for the specified lag period. What this means is that predictions are good on average, and there is only a fair-coin's worth of deviation between prediction and actual behavior.

Most importantly, the linear model does not permit prediction errors with temporal structure deviating from white noise, regressing-to-the-mean process. When prediction errors begin to deviate from this uncorrelated random variation, it signifies a systematic departure of the linear model's measurement series. Just isolated points in time do not tip us off to something being amiss. It is a statistical symptom that will only show up as we examine how the linear model compares to the time series in the long run. That departure might be sudden and abrupt; it may be continuous or intermittent. However, in the long run, the issue is that the prediction errors could be correlated with themselves across time—across those same lags we had seen in the autocorrelation function above. This departure of prediction errors from white noise is the empirical margin within which multifractal methods might help.

**How do we perform multifractal analysis?**

The present work uses one of the most straightforward variants of multifractal analysis built on the foundation of 'bin proportions' (Fig. 3). We will unpack this idea of bin proportion as follows: 'bins' stand for subsections of the measured time series and can also be called 'time windows,' 'limited samples' or 'short snippets' of the longer time series. The question of concern is how closely any single bin of the time series (i.e., any small subset of measurements over time) resembles other measurements over time. For example, will one bin of the time series look like another bin at some other time in the same measurement? How do these subsets of measurements vary when looking at different timescales—do measurements in one bin look like the measurements found in a longer bin?

We will use the mathematical language of 'bin proportion'—fractions to express probabilities—to compare different bins. Proportion indicates a small subset we observe divided by the entire set. Bin proportion is obtained when "the amount of the measured stuff in one bin" is divided by "the amount of all measured stuff across the whole time series."

The unifying mathematical question fundamental to multifractal analysis is: "how does proportion change with time scale?" This question can be placed in terms of the daily rainfall example. As rainfall is observed over many days, the amount of rainfall can be compared to the time length of measurement. For instance, let us imagine that an impatient observer measures rainfall only for two days; if it rains the same amount two days in a row, then each day represents half of the total rainfall over two days. However, if it rains three times as much on the first day as it rains on the second day, then the first day's rainfall represents 75% of the total rainfall over two days. Let us say that a more patient observer measures rainfall for ten days. In this case, if it rains the same amount every day, then each day's rainfall represents 10% of the total rainfall over ten days.

Let us consider how these outcomes align with our examples of possible outcomes for our time series. If we assume the simplest case of perfect stability (e.g., perfectly uniform amounts of rainfall across days), then we should predict that each day represents the same proportion of all the total rainfall over multiple days and that the proportion of the rainfall represented by a single day will decline with the increasing number of days. Mathematically, we find the following relationship:

Each day's proportion of uniform rainfall = 100% of total observed rainfall / Number of days observed

This relationship is a power law where the day's proportion $P$ is an inverse power law of the number of days observed $N$, that is, $P = 1/N$ or $P = N^{-1}$. This relationship is estimated on logarithmically scaled axes because:

$$\log P = -1 \times \log N. \tag{1}$$

In Eq. (1) The number multiplied by $\log N$ is called the singularity strength. Here, 'singularity' indicates that the power-law exhibits a single form across all scales. That is, no matter how many more days a perfectly uniformly distributed process is observed, this same relationship holds.

**Perfectly uniform changes over time yield zero multifractality**

Just as white noise converges to a stable standard deviation in the long run, the same singular relationship will be seen when considering longer bins (i.e., greater amounts of time and so longer subsets of the time series). For instance, pooling the rainfall over non-overlapping 4 days (days 1 through 4 in one bin, days 5 through 8 in another bin, days 9 through 12 in yet another bin, and so on…) allows examining rainfall in 4-day increments. And for a very long time series, the average 4-day proportion of total rainfall will converge towards the $P = N^{-1}$ relationship noted above.

Now, we explore the concept of singularity strength developed in the previous section with regard to the rainfall example. The singularity strength is a power-law exponent that describes explicitly how bin proportions grow with longer bins. In other terms, if only one short subset of our time-series measurements is considered, the singularity strength tells us how representative any given subset is of the entire measurement. So, stronger singularity strength means that shorter subsets of the observations provide more clues about the extent of that measure across the whole time series. For our rainfall example, stronger singularity strength suggests that short glimpses of rainfall over a few days may be more informative about the longer-term rainfall over the years.

Something important to remember is that multifractality is entirely about the number of singularity strengths that can be estimated from the measured time series. The most straightforward measure of multifractality is the range of the singularity strengths (i.e., maximum minus minimum). So, for perfectly uniform rainfall and homogeneous white-noise rainfall, multifractality is zero. In both cases, the bin proportion of the rainfall is a single stable function of the observation timescale.

If rainfall is perfectly uniform or white noise, then at any bin size—any time scale—the proportion of the total rainfall found in a random bin of that size can be predicted. For instance, looking at either half of the time series allows seeing half of the total rainfall; looking at any quarter of the time series allows seeing a quarter of the total rainfall. A one-to-one relationship exists between the proportion of time and the measurement proportion for all timescales. So, these timescales can be characterized by the same singularity strength: one and only one power-law relationship exists between bin proportion and bin size (i.e., Bin proportion ~ Bin size). The fact that only one power law describes this relationship (i.e., with exponent = 1) means that only one singularity strength exists, and so the range of singularity strengths is zero, entailing an absence of multifractality.

The word 'multifractality' is a little more opaque than necessary—Mandelbrot (2013) himself recognized as much and complained that other scholars had renamed his modeling strategy poorly. This term is cluttered with the 'fractal' root, indicating a fractional singularity strength (this term 'fractional' is discussed in this essay for the first time now because the only singularity strength seen above so far is equal to 1). Saying that singularity strengths can be fractional means that, whereas Bin Proportion ~ Bin size$^1$, it is equally possible to observe time series for which Bin Proportion ~ Bin size$^{1.3}$ or Bin Proportion ~ Bin size$^{0.6}$. So, for multiple reasonable mathematical possibilities, mean and standard deviation can suffice because, just like linear autocorrelation or the Fourier transform, multifractality can be zero-valued. But for most empirical time series, the singularity strength is a nonzero number.

**Irregular changes over time entail greater-than-zero multifractality**

Multifractality results from the measured time series displaying irregular temporal sequence. This point means that, suddenly, the measured process develops over time in a way that does not resemble past changes over time. That is, multifractality grows beyond zero, indexing temporal asymmetry. To put multifractality in contrast with more familiar time-series structure, white-noise epitomizes time symmetry. So, multifractality contrasts with white noise.

The changes exhibiting irregularity that registers multifractality (i.e., nonzero range of singularity strength) can be conceived in two ways, and both yield mathematically the same results. The first option is to think about irregularity as changes in singularity strength over time. A rough way to characterize this option might be that, for an irregular time series, the first half of the series exhibits one power-law $P = N^{-a}$, and the second half exhibits another power-law $P = N^{-b}$ such that a ≠ b. When examining white-noise rainfall, these halves would each contain the same proportion of the total rainfall. The second option is that more rainfall days show one power-law $P = N^{-c}$, and less rainfall days show another power-law $P = N^{-d}$, such that c ≠ d. When examining white-noise rainfall, the days with less rainfall would show the same power-law as the days with more rainfall. In both options, multifractality would be the nonzero absolute difference between a and b or between c and d. In these brief examples of how time series could change irregularly, we do not suggest any reliable relationship between sequence and size. For instance, there would be no cyclicity, and neither are we suggesting that the first half of the series contains all of the larger daily rainfall values (i.e., that a = c). The goal is to only illustrate that singularity strengths can vary with the measurements' time and magnitude. The next section will address how we estimate this from variation.

Despite the conceptual access to multifractality through these two modes (i.e., "change with time" or "change with magnitude"), many multifractal analyses assume the latter mode as the default. That is, most multifractal analyses proceed by evaluating singularity strengths for measurements of different sizes as opposed to across different time windows. A good reason for this strategy is that evaluating changes in singularity strength with time is often puzzling. For instance, in the examples above, the singularity strength was illustrated in perfectly uniform time series, and to do so, how proportions were spread across the whole time series of rainfall was considered. To examine how singularity strength changes with time, the measured time series had to be chopped up into many smaller time series. There is nothing wrong with this approach, and it can be quite useful (Grech and Pamuła 2008), but there are some unresolved questions about how many subsets to estimate and what length of subsets are going to provide informative estimates (França et al. 2018). The good news is that various attempts converge on a similar or even identical estimate of multifractality as the "change with magnitude" method does, so it is not a question of accuracy (Ihlen 2012). Both methods are just as effective and easily used. The "change with time" variant of multifractal analysis is only slightly less common than the "change with magnitude" variant.

**Estimating multifractality in terms of how the singularity strength changes with magnitude**

The "change with magnitude" calculation of multifractality is a mathematically compact way to proceed. It might seem like evaluating singularity strengths for 'larger' measurements (e.g., for days with more rainfall) or for 'smaller' measurements (e.g., for days for less rainfall) could be as slippery as the question of how often to check for changes in the singularity strength over time—for instance, should 'larger' vs. 'smaller be defined?' These distinctions sound challenging to define and can be no more likely to be correctly guessed than to identify all appropriate subsets to estimate in the "change with time" strategy above.

Fortunately, the multifractal formalism manages to generalize the 'larger' vs. 'smaller' distinction into a more continuous framework by spreading this dichotomy across several degrees of relative sizes. This strategy bypasses the need to rely on any single cutoff for magnitude to determine the singularity strength. This feat is accomplished using an exponent abbreviated as '$q$,' a parameter that can be set to any real value (Fig. 4). There is no correct value of $q$, and more values of $q$ could potentially reveal new singularity strengths. The point is not to uncover "all of the multifractality" with "all of the right $q$ values." The point is to demonstrate that there is a variation of singularity strength with any values of $q$, and no difference in singularity strengths is found for a given $q$, that failure to find multifractality is itself informative about the regularity of the temporal sequence. Whatever value of $q$ is used, it is applied to the proportion estimated for any given bin. When using this proportion-based multifractal analysis, a q exponent equal to 1 leaves the proportions unchanged. For instance, if we are looking at a 4-day bin that contains 1/10 of the total rainfall, then $q = 1$ gives us the same value.

When $q$ is set to values different from 1, it systematically emphasizes different parts of the same time series. Different values of $q$ will systematically increase or decrease differently-sized bin proportions. By applying a $q$ other than 1, we reshape the sequence of bin proportions, producing a different power-law relationship between bin proportion and bin size. In simple terms, gradually increasing $q$ allows statistically zooming in and zooming out fluctuations of various sizes, for instance, days or groups of days with relatively more or less rainfall. Hence, with each reshaping of the time series and each value of $q$, we can estimate the singularity strength. For a perfectly uniform time series, as we have stressed above, there is no multifractality. So, applying these different values of $q$ will most certainly not have any effect on the estimated singularity strength. However, manipulating the $q$ exponent can show us different singularity strengths within the same series when changes over time have been irregular. This variety of singularity strengths may indicate an irregularity beyond linear modeling and more suggestive of nonlinearity.

How the math of $q$ parameter works can be seen with a very small example beyond $q$ equaling 1. As $q$ increases beyond 1, bigger $q$ makes proportions smaller. For instance, applying a $q$ of 2 to the 1/10 proportion at the 4-day bin that contains 1/10 of the total rainfall generates a new and much smaller number 1/100. Although all exponents bigger than 1 will make any fraction smaller, $q$ greater than 1 will have a more drastic effect on smaller fractions. For instance, 1/4 is greater than 1/10, and applying a $q$ of 2 to each of them yields $(1/4)^2 = 1/16$ and $(1/10)^2 = 1/100$, respectively. When $q$ equals zero, the bin proportions measured as 1/4 and 1/10 both collapse to the same value: $(1/4)^0 = 1$ and $(1/10)^0 = 1$, respectively. However, when $q$ is negative, the ordering of these proportions reverse—the bigger proportion becomes the smaller of the two, and vice versa: $1/4^{-1} = 4$ and $(1/10)^{-1} = 1$, respectively.

Reshaping bin proportions profile across the measured time series with the $q$ exponent unearths the variety and nonlinearity of temporal structure in a time series. Critically, applying $q$ does not affect the estimated singularity strengths in a uniform sequence like white-noise time series. The multifractal analysis is the estimation of singularity strengths for a variety of values of $q$; the procedure includes a few statistical checks to ensure the stability of these estimates of singularity strengths. However, the fact that not all singularity strengths are equal is evidence that the time series is 'multifractal,' and the range (i.e., maximum minus minimum) of singularity strengths is the simplest estimate of 'multifractality' as opposed to the comparison of time-dependent bin. Further details on calculation and interpretation can be found in tutorials from elsewhere (Ihlen 2012; Kelty-Stephen et al. 2013).

**How to determine when multifractality implies that the time series is nonlinear?**

The link between multifractality and nonlinearity is subtler than what many audiences may expect. Rhetorically, multifractality has become firmly linked to discussions of nonlinear dynamics, especially because establishing nonlinear interactions across scales requires a specific kind of multifractal evidence (e.g., Ihlen and Vereijken 2010). However, multifractality does not necessarily imply nonlinearity, and very many linear processes can exhibit a multifractal structure. Instead, the clearer, more consistent connection is between multifractality and irregularity across time. As we have noted above, irregularity can be linear, and nonlinear models need to be invoked only when the irregularity indicates prediction errors failing to have a white-noise structure. On the face of it, the multifractality of a time series has nothing especially informative by itself except for irregularity across time. When it increases beyond zero, it simply means that the time series is not perfectly uniform white noise.

Multifractality can appear in linear changes over time because some irregularity across time can be linear (e.g., in the case of cyclical changes as rainfall during specific months every year). As already pointed out above, many measured time series can have irregular changes over time, but that does not prevent them from being linearly modellable and showing trends of regression to the mean. The question of whether multifractality can appear in linear systems concerns those prediction errors noted above. So long as the prediction error series is devoid of temporal structure as white noise, the process is likely linear—Fourier-transform-based simulation is used to test this idea, as elaborated below. Similarly, a wide margin of error is expected from linear sources that could cause multifractal indications of irregularity. Indeed, perfectly linear changes can be multifractal. All that is needed to demonstrate linearity is for white-noise errors to separate a measured irregular time series from its best-fitting linear model.

**How to determine when multifractality reflects nonlinear interactions across scales?**

The question essentially boils down to whether multifractality in the measured time series is different from what the best-fitting linear model suggests. We estimate multifractality for the original time series to address this question, and we estimate multifractality for a finite set of time series built to mimic linear properties of the original time series. A procedure called the iterative amplitude adjusted Fourier transform (IAAFT) is outlined in what follows below. This procedure was developed by Schreiber and Schmitz (1996) to test for nonlinearity and recommended by Ihlen and Vereijken (2012) to test whether multifractal results reflect nonlinear interactions across scales. Ultimately, the way to determine when multifractality reflects nonlinear interactions across scales is when the original series' multifractality is significantly different from the finite set of IAAFT series' multifractality.

**Sketch of the IAAFT procedure**

Fig. 5 provides a step-by-step illustration of the Iterative Amplitude Adjusted Fourier Transform (IAAFT) procedure. The "best-fitting linear model" for changes over time appears as the Fourier transform's amplitude spectrum. As noted above, the Fourier transform is a mathematical description of a time series that uses sinusoidal waves to describe a time series, and the amplitude spectrum is what is called the list of amplitudes corresponding to each possible frequency of sinusoidal waves. It was noted above that the Fourier transform's amplitude spectrum bears a one-to-one relationship with the linear autocorrelation.

 Testing whether multifractality in the measured time series is due to nonlinear interactions across scales requires representing how much variability in multifractality would be typical of a linear model of changes over time. Hence, while the amplitude spectrum of the original time series has to be used, the multifractality of another time series with the same amplitude spectrum must also be estimated. So, all we know without further test is the estimate of multifractality for the original series. If we want to know if multifractality reflects nonlinearity, then we need to know what amount of multifractality the linear model would predict. To learn that, then we need to estimate multifractality with a new time series that has the same linear model as our original series. What is needed is to generate a new time series with the same linear temporal structure and even the same mean and standard deviation, but it is needed to manifest in a different time series than the one measured.

 Understanding this next step requires stepping aside to unpack another detail of the Fourier transform. The Fourier transform also estimates a phase spectrum. The details of the phase spectrum are not necessary, except that it determines the sequence in which each of the sinusoidal waves appears in the linear model. The amplitude spectra tell us how large sinusoidal waves composing a time series are, and phase spectra tell us how far separated in time each frequency's corresponding sinusoidal waves are. Critically, the linear model assumes independence across time, and hence, the phase spectrum is not essential to the linear model; it is just a byproduct of statistical accounting. So, we can make a linear version of the measured time series simply by applying the inverse Fourier transform to two elements: 1) the original series' amplitude spectrum, and 2) a random reshuffling of the phase spectrum (i.e., randomly assigning the same phases to different frequencies). In a sense, the waves found in the original times series are chopped and put in a different sequence. The phase spectrum is at the bottom of our original question: whether changes over time differ across the beginning, middle, and end of a developing process. Scrambling the phase spectrum breaks the originally measured sequence and tests whether the phase ordering matters. Scrambling the phase spectrum while keeping the amplitude spectrum unchanged gives a linearly equivalent time series with the original time series' sinusoidal rises and falls. This step allows reaching the heart of the contrast between original series' multifractality and corresponding IAAFT series' multifractality.

 At this point, an inverse Fourier transform's values follow a generic Gaussian distribution with a mean equal to zero and a standard deviation determined by default settings on a computer program for the Fourier transform. The next step in the procedure is to rank-order the values in the computer-generated inverse Fourier series and rank-order the values in the measured time series. The inverse Fourier series values are then replaced with rank-matched values of the measured time series. That is, the maximum value of the inverse-Fourier series is replaced by the maximum value of the measured series, the second-highest value of the inverse-Fourier series is replaced by the second-highest value that the measured series replaces, and so on…. Replacing all of these rank-matched values allows maintaining the measured series' mean and standard deviation while preserving some of the original amplitude spectra. All of this replacement is done by a computer program, and the result is a series of the same values with the same amplitude spectrum but none of the sequence of the original time series.

However, because the histogram of the measured time series is non-Gaussian (i.e., it is skewed or contains kurtosis), rank-matching may distort the sinusoidal forms entailed by the amplitude spectrum. So, the amplitudes need adjusting, the 'AA' in IAAFT. This adjustment is accomplished by iteration: replacing the originally measured values by rank matching distorts the amplitude, repeating the process. The Fourier transform is taken of the new series whose amplitude spectrum has been deformed by rank matching. This process generates a slightly deformed amplitude spectrum and a new phase spectrum. The deformed amplitude spectrum is replaced by the original series' amplitude spectrum, and the new phase spectrum is scrambled. Then, the inverse-Fourier transform on this new scrambled phase spectrum and the old amplitude spectrum is repeated. This process reinstates the original, intact amplitude spectrum to ensure that rank-matching to replace the values does not weaken linear features compared to the original time series. Iteration allows breaking the original sequence continually and, each time, reinstating the original linear structure from the original series' amplitude spectrum.

**What is multifractal nonlinearity as described in the main text?**

What is called 'multifractal nonlinearity' in the main text above is a one-sample $t$-test comparing the original series' multifractality to multifractality of a finite set of IAAFT surrogates with matching linear structure (Fig. 6). As the original series' multifractality departs more and more from the multifractality attributable to the linear structure of IAAFT surrogates, this $t$-statistic grows larger. If the $t$-statistic in this comparison is significantly large (i.e., $p < 0.05$), then the irregularity of the measured time series is interpreted as so strong that it requires a nonlinear model. Fig. 7 provides examples of multifractal analysis to quantify the strength of multifractal nonlinearity in a White noise series, and postural center of pressure (CoP) displacement series of a person standing quietly with the eyes fixated at a distant point [data from Mangalam et al. (2021)].

**Discussion**

Nonlinearity can sometimes inflame passions and generate more conflict than progress (e.g., Van Orden et al. 2003; Wagenmakers et al. 2004, 2012; Dixon et al. 2012). However, all of the preceding discussion offers a roadmap to approach nonlinearity with a sober view of the possibilities. For instance, the mounting evidence that nonlinearity matters to psychological experience is intriguing and worth further and broader attention (Kelty-Stephen et al. 2016, 2020, 2021; Teng et al. 2016; Carver et al. 2017; Ward and Kelty-Stephen 2018; Doyon et al. 2019; Mangalam and Kelty-Stephen 2020; Jacobson et al. 2020; Mangalam et al. 2020; Bloomfield et al. 2021). We hope that multifractality in our measures speaks directly to the capacity of developing systems to blend processes across different timescales, producing emergent structures (e.g., Gottlieb 2002). It remains for a larger discussion across the field to give this question adequate treatment.

The preceding discussion provides a constructive lens through which this future work can proceed. We must be clear about what the groundswell of multifractality does mean and what it does not mean. Multifractality encodes a change that the linear model cannot, but the bare prevalence of multifractal results does not imply nonlinearity. Multifractality is not an invitation to eschew all linear modeling but instead to take closer stock of what linear modeling can and cannot accomplish. Additionally, multifractality explicitly models those portions of measurement for which linearity fails to apply. Indeed, not all that is multifractal is necessarily nonlinear, and multifractality does not replace linear modeling of the linear aspects of measurements. The comparison to surrogates offers a subtle tempering of our interest in nonlinearity by drawing into relief precisely how strong the nonlinearity is; that is, we offer here a way to judiciously take both linear and nonlinear aspects of measured change into consideration.

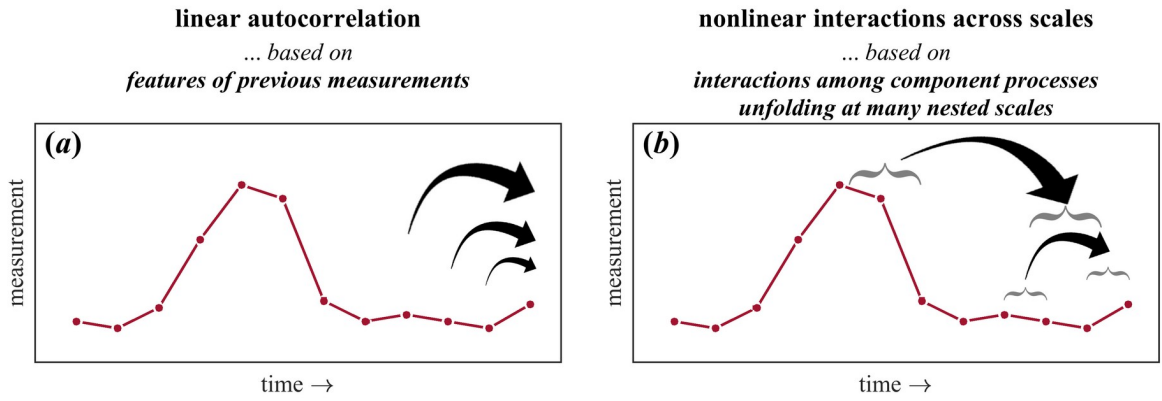

**Fig. 1.** Two perspectives about how a measured time series is analyzed. (*a*) The linear autoregressive perspective takes the premise that each measure in time entails the summing of random and independent factors. (*b*) The multifractal perspective takes the premise that each measure entails interactions among component processes at many nested scales.

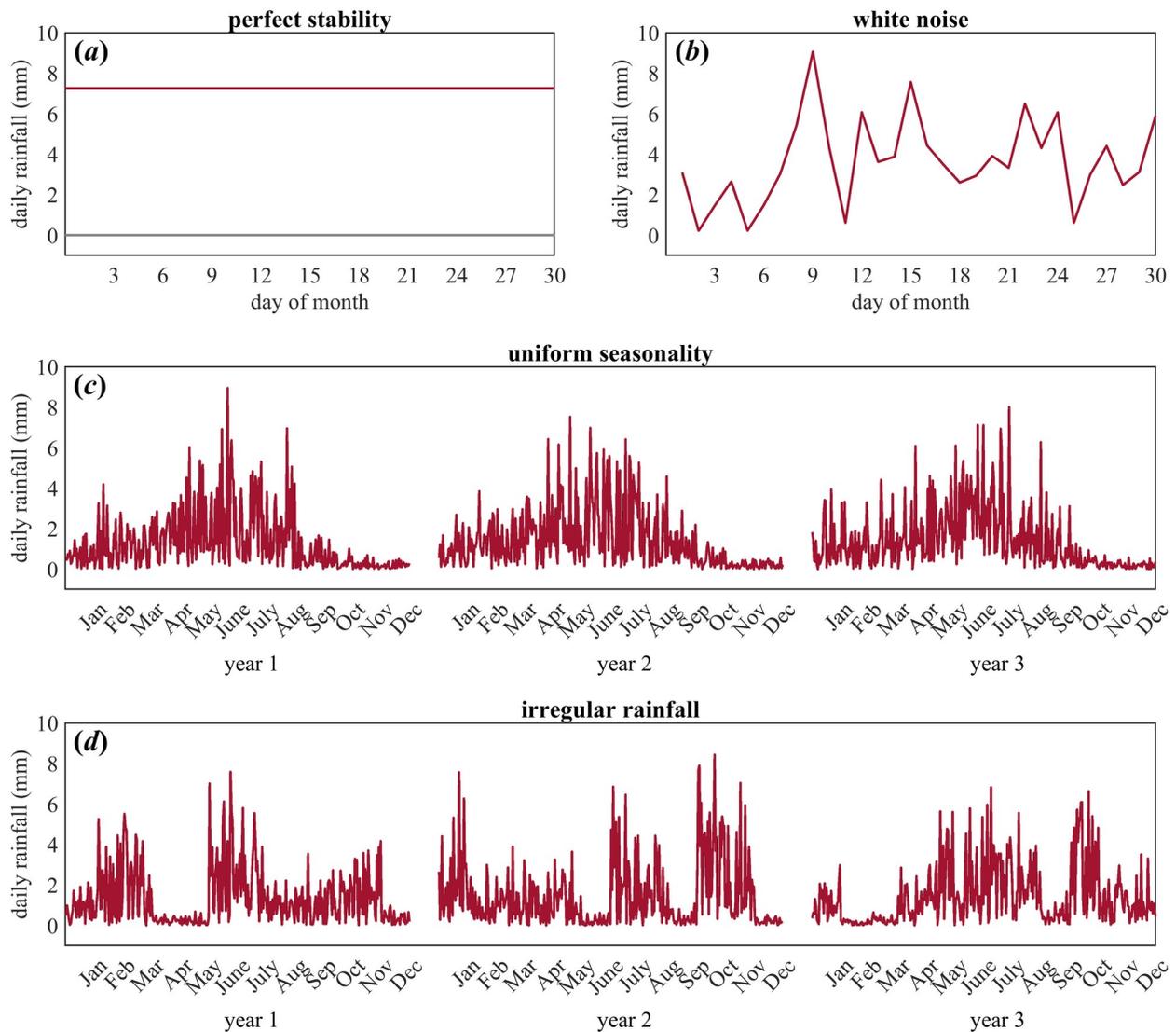

**Fig. 2.** Possible time series of daily rainfall in a given region. (*a*) Perfect stability. The simplest measurements of daily rainfall include no rainfall (gray line) or always the same amount of rainfall (red line). (*b*) White noise. Uncorrelated random variation of daily rainfall, varying according to white noise. (*c*) Uniform seasonality. The measurement area may have a rainy season and, therefore, a cyclical rainfall pattern (e.g., more in June and July than in April). (*d*) Irregular seasonality. Rainy seasons can come late one year or early; they can come early multiple years and late the following year. Also, not shown here, but wet or dry years can exist in which the rainy season varies in intensity across years, decades, and centuries.

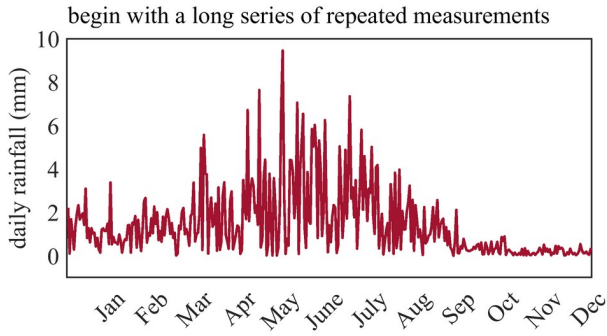

begin with a long series of repeated measurements

bin measurement with different window sizes, **L**

**larger bins contain**, on average, **greater proportions** of the sum of all measurements, $P$

linear slope of **log $L$** vs. **log$P(L)$** yields **singularity strength, $\alpha$**

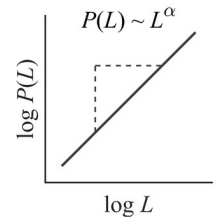

$P(L) \sim L^\alpha$

**smaller bins contain**, on average, **greater *variability* in proportions** of the sum of all measurements, $P$

slope from linear regression yields **Hausdorff dimension, $f$**

$L\ P$ *Entropy*

$L\ P$ *Entropy*

$L\ P$ *Entropy*

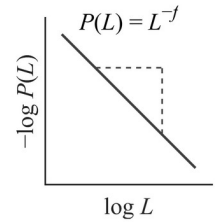

$P(L) = L^{-f}$

**Fig. 3.** Multifractal analysis is founded on 'bin proportions.' 'Bins' stand for subsections of the measured time series. Bin proportion is obtained when "the amount of the measured stuff in one bin" is divided by "the amount of all measured stuff across the whole time series." See section "How do we perform multifractal analysis?" for details.

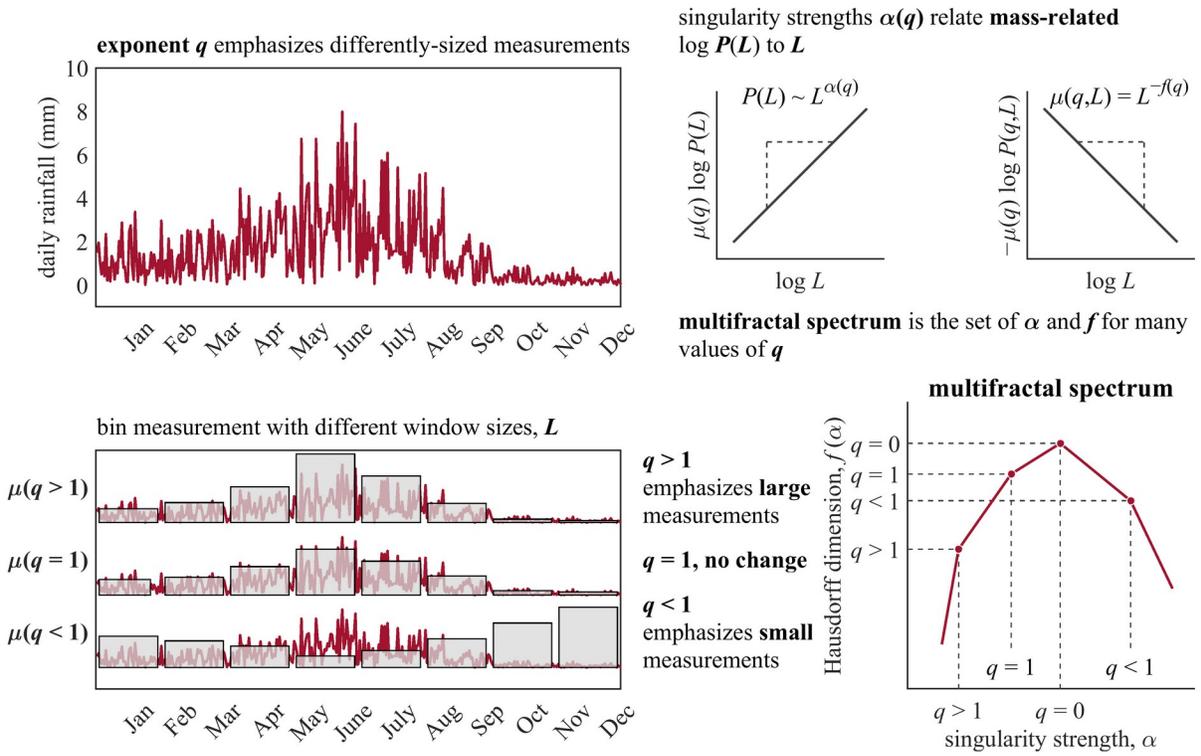

**Fig. 4.** Exponent *q*. When *q* is set to values different from 1, it systematically emphasizes different parts of the measured time series. As *q* increases beyond 1, bigger *q* makes bin proportions smaller. Although all $q > 1$ will make any fraction smaller, *q* greater than 1 will have a more drastic effect on smaller fractions. For instance, 1/4 is greater than 1/10, and applying a *q* of 2 to each of them yields $(1/4)^2 = 1/16$ and $(1/10)^2 = 1/100$, respectively. When *q* equals zero, the bin proportions measured as 1/4 and 1/10 both collapse to the same value: $(1/4)^0 = 1$ and $(1/10)^0 = 1$, respectively. However, when *q* is negative, the ordering of these proportions reverse—the bigger proportion becomes the smaller of the two, and vice versa: $1/4^{-1} = 4$ and $(1/10)^{-1} = 1$, respectively. See section "Estimating multifractality in terms of how the singularity strength changes with magnitude" for details.

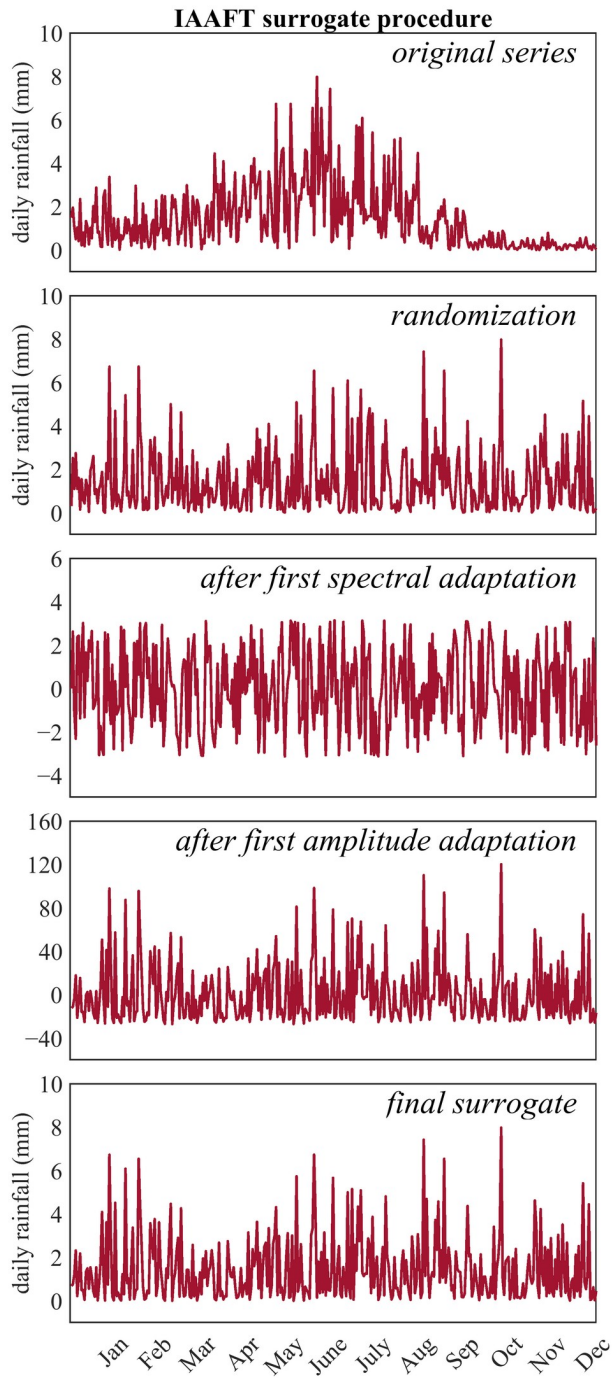

**Fig. 5.** An illustration of the Iterative Amplitude Adjusted Fourier Transform (IAAFT) procedure using the dummy data for daily rainfall. The steps of spectral adaptation and amplitude adaptation are iterated until the original, intact amplitude spectrum is reinstated. See section "Sketch of the IAAFT procedure" for details.

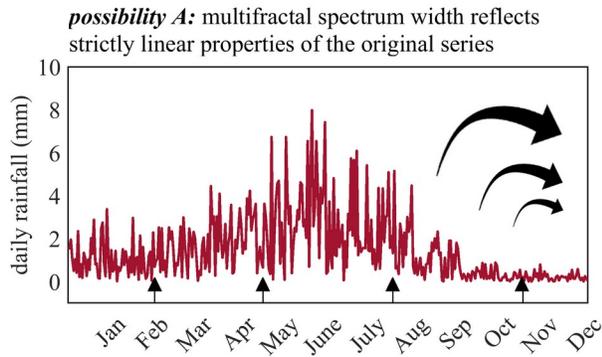
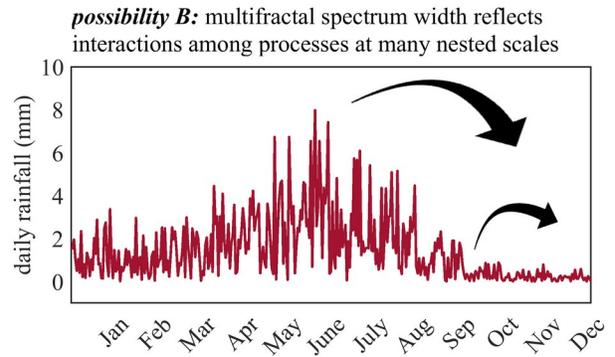
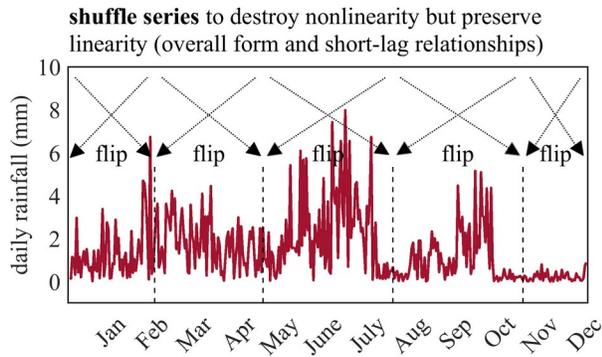

**Fig. 6.** IAAFT surrogate analysis used to identify whether the measured time series exhibits multifractal nonlinearity. See section "What is multifractal nonlinearity as described in the main text?" for details.

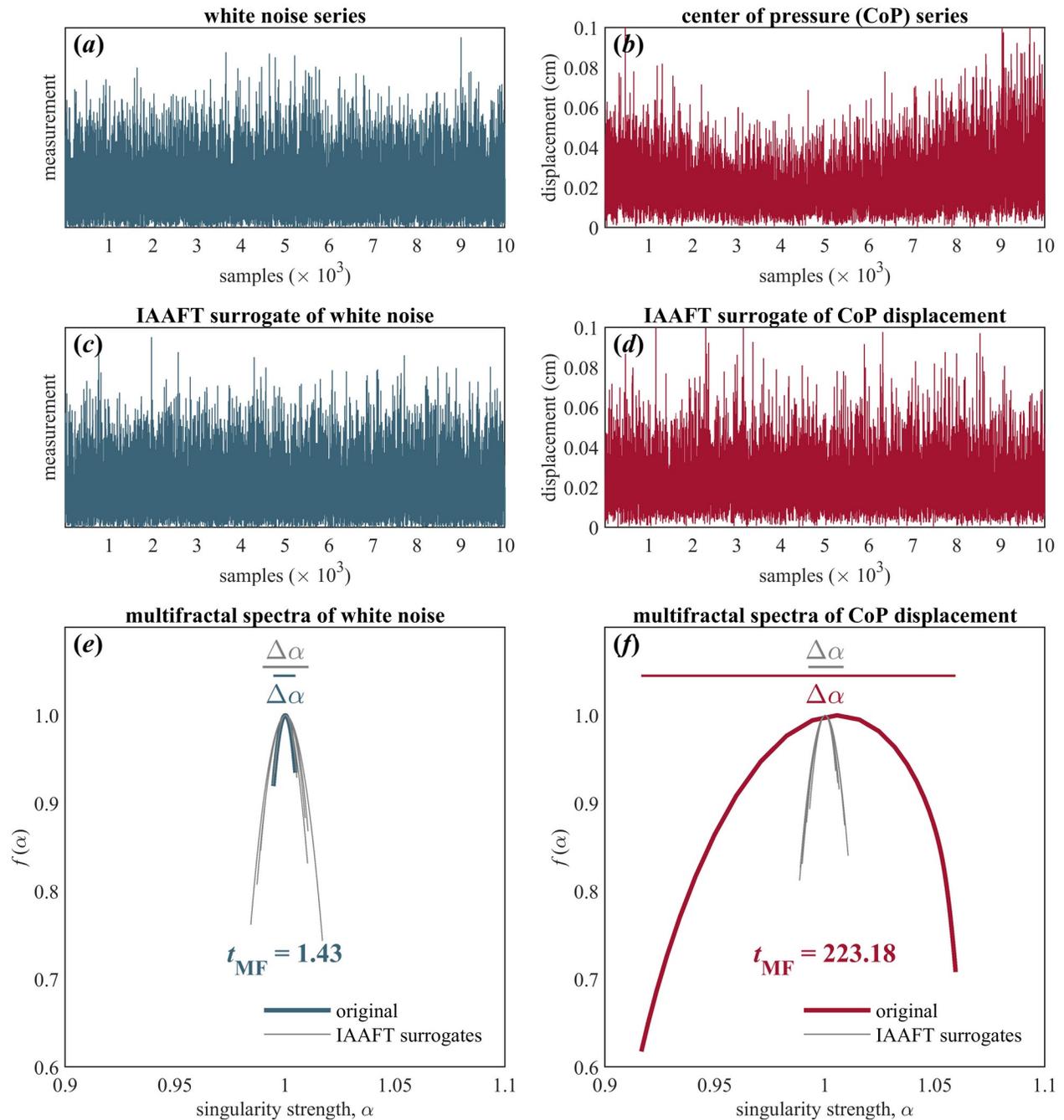

**Fig. 7.** Multifractal analysis to quantify the strength of multifractal nonlinearity. (*a*) White noise series. (*b*) Postural center of pressure (CoP) displacement series of a person standing quietly with the eyes fixated at a distant point [data from Mangalam et al. (2021)]. (*c*) IAAFT surrogate of white noise in (*a*). (*d*) IAAFT surrogate of CoP displacement. (*e*) No significant difference in multifractal spectrum width of white noise series and multifractal spectra widths of its IAAFT surrogates ($n = 32$) suggests an absence of multifractal nonlinearity. Only five IAAFT spectra are shown for clarity. (*f*) Significantly larger multifractal spectrum width of CoP displacement series than multifractal spectra widths of its IAAFT surrogates ($n = 32$) suggests multifractal nonlinearity. Only five IAAFT spectra are shown for clarity.

## APPENDIX A

### Direct estimation of multifractal spectra using Chhabra and Jensen's method

The multifractal spectrum is a relationship between two fractal dimensions $\alpha$ and $f(\alpha)$. This Chhabra and Jensen's (1989) method samples non-negative series $u(t)$ at progressively larger scales such that proportion of signal $P_i(L)$ falling within the $i^{th}$ bin of scale $L$ is

$$P_i(L) = \frac{\sum_{k=(i-1)L+1}^{iL} u(k)}{\sum u(t)} \qquad (1)$$

As $L$ increases, $P_i(L)$ represents progressively larger proportion of $u(t)$,

$$P(L) \propto L^{\alpha} \qquad (2)$$

suggesting growth of proportion according to one 'singularity' strength $\alpha$. $P(L)$ exhibits multifractal dynamics when it grows heterogeneously across timescales $L$ according to multiple potential fractional singularity strengths, such that

$$P_i(L) \propto L^{\alpha_i} \qquad (3)$$

whereby each $i^{th}$ bin may show a distinct relationship of $P(L)$ with $L$. The spectrum of singularities is itself the multifractal spectrum, and its width $\Delta\alpha$ ($\alpha_{max} - \alpha_{min}$) indicates the heterogeneity of these relationships between proportion and timescale (Halsey et al. 1986; Mandelbrot 1999).

This method estimates $P(L)$ for $N_L$ nonoverlapping bins of $L$-sizes and accentuates higher or lower $P(L)$ by using parameter $q > 1$ and $q < 1$, respectively, as follows

$$\mu_i(q, L) = \frac{[P_i(L)]^q}{\sum_{i=1}^{N_L} [P_i(L)]^q} \qquad . \qquad (4)$$

See Fig. 2.

$\alpha(q)$ is the singularity for $\mu(q)$-weighted $P(L)$ estimated by

$$\alpha(q) = -\lim_{L \to \infty} \frac{1}{\ln L} \sum_{i=1}^{N} \mu_i(q, L) \ln P_i(L)$$

$$= \lim_{L \to 0} \frac{1}{\ln L} \sum_{i=1}^{N} \mu_i(q, L) \ln P_i(L) \qquad . \qquad (5)$$

Estimates $\alpha(q)$ belong to the multifractal spectrum if Shannon entropy of $\mu(q, l)$ scales with $L$ according to a dimension $f(q)$, where

$$f(q) = -\lim_{L \to \infty} \frac{1}{\ln L} \sum_{i=1}^{N} \mu_i(q, L) \ln \mu_i(q, L) \qquad . \qquad (6)$$

For $q$ generating scaling relationships (Eqs. 5 and 6) with correlation coefficient, $r > 0.95$, the parametric curve $(\alpha(q), f(q))$ or $(\alpha, f(\alpha))$ constitutes the multifractal spectrum with width $\Delta\alpha = \alpha_{max} - \alpha_{min}$.

Critically, higher $r$ is better. To the degree that researchers become interested in comparison to surrogates, it can be useful to reduce this correlation benchmark. The reason is that if all we are interested in is multifractal-spectrum width, then it may be appropriate to leave the estimate of the multifractal-spectrum width at zero. However, this conservativism can backfire when also calculating multifractal-spectrum widths for surrogates which usually have more stable scaling relationships. Hence, a zero-width multifractal spectrum for an original series can co-occur with non-zero multifractal spectrum widths for the surrogates.

# APPENDIX B

## R code for Chhabra and Jensen's (1989) method

```r
minq <- -10 #minimum q that you want to test
maxq <- 10 #maximum q that you want to test
qstep <-
  1 #increment that you want to use to test values between the above min
and max
#data should be a column vector
chhabraJensen <- function(data, minq, maxq, qstep) {
  minbin <- 4
  ldat <- length(data)
  maxbin <- floor(ldat / 4)
  qvec <- seq(minq, maxq, by = qstep)
  Fqvalsorig <- mat.or.vec((maxq - minq) / qstep + 1, 1)
  Fqcorrorig <- mat.or.vec((maxq - minq) / qstep + 1, 1)
  Aqvalsorig <- mat.or.vec((maxq - minq) / qstep + 1, 1)
  Aqcorrorig <- mat.or.vec((maxq - minq) / qstep + 1, 1)
  lqvec <- length(qvec)
  bins <- mat.or.vec(maxbin - 3, 1)
  Fqnumer <- mat.or.vec(maxbin - 3, lqvec)
  Aqnumer <- mat.or.vec(maxbin - 3, lqvec)
  ##commented-out lines can replace the prior three lines
  #binsizes <- 2^c(2:floor(log(maxbin)/log(2)))
  #bins <- mat.or.vec(length(binsizes),1)+binsizes
  #Fqnumer <- mat.or.vec(length(binsizes),lqvec)
  #Aqnumer <- mat.or.vec(length(binsizes),lqvec)

  for (currinc in 4:maxbin) {
    bins[currinc - 3] <- currinc
    #for (count in 1:length(bins)){
    #currinc <- bins[count]
    incrementBins <- floor(length(data) / currinc)
    datacurr <- data[1:(floor(length(data) / currinc) * currinc)]
    dataMat <- matrix(datacurr, nrow = currinc, byrow = FALSE)
    Tivec <- colMeans(dataMat)

    sTiv <- sum(Tivec)
    Pivec <- Tivec / sTiv
    for (qcount in 1:lqvec) {
      qq <- qvec[qcount]
      PNZq <- Pivec ^ qq
      SPNZq <- sum(PNZq)
      Mivec <- (PNZq) / SPNZq
      lM <- log(Mivec)
      entMbit <- Mivec * lM
      entM <- sum(entMbit)
      Fqnumer[currinc - 3, qcount] <- entM
      PVC <- Pivec
      lPVC <- log(PVC)
      MlPVC <- Mivec * lPVC
      sMlPVC <- sum(MlPVC)
      Aqnumer[currinc - 3, qcount] <- sMlPVC
    }
```

```
    }
    critAs <- 0

    for (qcount in 1:lqvec) {
      binsqF <- bins
      Fqnumerq <- Fqnumer[, qcount]
      Fqnumerq <- Fqnumerq[!is.na(Fqnumerq)]
      Fqnumerq <- Fqnumerq[is.finite(Fqnumerq)]
      ldiffF <- length(Fqnumer[, qcount]) - length(Fqnumerq)
      binsqF <- binsqF[(1 + ldiffF):length(binsqF)]
      binsqA <- bins
      Aqnumerq <- Aqnumer[, qcount]
      Aqnumerq <- Aqnumerq[!is.na(Aqnumerq)]
      Aqnumerq <- Aqnumerq[is.finite(Aqnumerq)]
      ldiffA <- length(Aqnumer[, qcount]) - length(Aqnumerq)
      binsqA <- binsqA[(1 + ldiffA):length(binsqA)]

      #####if you want to increase speed, you may use these #-prefaced lines in place of the lines that #follow
      #if (length(binsqF)==length(Fqnumerq) && length(binsqA)==length(Aqnumerq) && length(Aqnumerq)>1
      #&&length(Fqnumerq)>1){
      #       if (cor(log(binsqF),Fqnumerq)>.95){ #.95 is the correlation used as the benchmark
      #               if (cor(log(binsqA),Aqnumerq)>.95){

      #                  slopeF <- .lm.fit(x=cbind(mat.or.vec(length(binsqF),1)+1,log(binsqF)),y=Fqnumerq)
      #                  slopeA <- .lm.fit(x=cbind(mat.or.vec(length(binsqA),1)+1,log(binsqA)),y=Aqnumerq)
      #                  Fqvalsorig[qcount]<-slopeF$coef[2]
      #                  Aqvalsorig[qcount]<-slopeA$coef[2]
      #                  critAs<-rbind(critAs,Aqvalsorig[qcount])
      #                  #print(critAs)
      #               }
      #       }
      #   }
      #}

      #if (length(critAs)>1){
      #     RangeA <- range(critAs[2:length(critAs)])
      #     RangeA <- RangeA[2]-RangeA[1]
      #} else {
      #    RangeA <- 0
      #}
      #return(RangeA)
      #}
      #####
      slopeF <- lsfit(log(binsqF), Fqnumerq)
      slopeA <- lsfit(log(binsqA), Aqnumerq)
      Fqvalsorig[qcount] <- slopeF$coef[2]
      Aqvalsorig[qcount] <- slopeA$coef[2]
      lbinsqA <- log(binsqA)
      lbinsqF <- log(binsqF)
```

```
      Fqcorrorig[qcount] <- cor(lbinsqA, Aqnumerq)
      Aqcorrorig[qcount] <- cor(lbinsqF, Fqnumerq)
    }

  critAs <- 0
  for (ii in 1:length(Aqvalsorig)) {
    if (Fqcorrorig[ii] > .95) {
      #.95 is the correlation used as the benchmark
      if (Aqcorrorig[ii] > .95) {
        critAs <- cbind(critAs, Aqvalsorig[ii])
      }
    }
  }
  RangeA <- range(critAs[2:length(critAs)])
  RangeA <- RangeA[2] - RangeA[1]
  return(RangeA)
}
```

# APPENDIX C

## R code for iterative amplitude adjusted Fourier transform (IAAFT)

```
library(gtools)
#data is time series
#max is maximum number of iterations, usually, 1000
iaaft2 <- function(data, copies, max) {
  ln <- length(data)
  fftdata <- fft(data)
  amp <- abs(fftdata)
  sgate <- matrix(rep(data, copies), ncol = copies)
  sgate <- apply(sgate, 2, sample)
  sortout <- sort(data, index.return = TRUE)
  fftsgate <- mvfft(sgate)
  fftsgate.im <- Im(fftsgate)
  fftsgate.re <- Re(fftsgate)
  fftrat <- fftsgate.im / fftsgate.re
  phase_x <- atan(fftrat)
  nn <- 1
  ind_prev <- sortout$ix
  sorted1 <- sortout$x
  while (nn <= max) {
    phase.i <- phase_x * 1i
    expphase <- exp(phase.i)
    sgate <- amp * expphase
    ifftsgate <- mvfft(sgate, inverse = TRUE)
    sgate <- Re(ifftsgate)
    lst <- apply(sgate, 2, sort, index.return = TRUE)
    sortstruct <-
      do.call(rbind, Map(cbind, colInd = seq_along(lst), lapply(lst, function(x)
        do.call(cbind, x))))
    ixMat <- matrix(sortstruct[, 3], ncol = copies)
    lst <- apply(ixMat, 2, sort, index.return = TRUE)
    sortstruct <-
      do.call(rbind, Map(cbind, colInd = seq_along(lst), lapply(lst, function(x)
        do.call(cbind, x))))
    ixMat <- matrix(sortstruct[, 3], ncol = copies)
    sgate <- matrix(sorted1[ixMat], ncol = copies)
    nn <- nn + 1
    fftsgate <- mvfft(sgate)
    fftsgate.im <- Im(fftsgate)
    fftsgate.re <- Re(fftsgate)
    fftrat <- fftsgate.im / fftsgate.re
    phase_x <- atan(fftrat)
  }
  return(sgate)
}
```